\documentclass[a4paper, conference]{IEEEtran}
\IEEEoverridecommandlockouts
\usepackage{cite}
\usepackage{amsmath,amssymb,amsfonts}
\usepackage{algorithmic}
\usepackage{graphicx}
\usepackage{textcomp}
\usepackage{xcolor}
\usepackage{pgfplots}
\usepackage{tikz}  
\usepackage{tikz-3dplot} 
\usetikzlibrary{shapes,arrows, positioning, automata, shadows,arrows.meta,backgrounds,fit, calc}
\usepackage{tkz-fct, tikz-dimline}
\usepackage{subcaption}
\pgfplotsset{compat=1.17} 
\usepackage{hyperref}

\newcommand\copyrighttext{%
  \footnotesize \textcopyright 2021 IEEE. Personal use of this material is permitted.
  Permission from IEEE must be obtained for all other uses, in any current or future
  media, including reprinting/republishing this material for advertising or promotional
  purposes, creating new collective works, for resale or redistribution to servers or
  lists, or reuse of any copyrighted component of this work in other works.
  DOI: \href{https://doi.org/10.1109/MMSP53017.2021.9733445}{10.1109/MMSP53017.2021.9733445}}
\newcommand\copyrightnotice{%
\begin{tikzpicture}[remember picture,overlay]
\node[anchor=south,yshift=10pt] at (current page.south) {\fbox{\parbox{\dimexpr\textwidth-\fboxsep-\fboxrule\relax}{\copyrighttext}}};
\end{tikzpicture}%
}

\tikzset{cross/.style={cross out, draw=black, minimum size=2*(#1-\pgflinewidth), inner sep=0pt, outer sep=0pt},cross/.default={4pt}}

\def\BibTeX{{\rm B\kern-.05em{\sc i\kern-.025em b}\kern-.08em
    T\kern-.1667em\lower.7ex\hbox{E}\kern-.125emX}}

\begin{document}

\title{Frequency-Selective Mesh-to-Mesh Resampling for Color Upsampling of Point Clouds\\
}

\author{\IEEEauthorblockN{Viktoria Heimann, Andreas Spruck and Andr\'e Kaup}
\IEEEauthorblockA{\textit{Multimedia Communications and Signal Processing}\\
\textit{Friedrich-Alexander University}, Erlangen-Nuremberg, Germany\\
\{viktoria.heimann, andreas.spruck, andre.kaup\}@fau.de}
}

\maketitle
\copyrightnotice

\begin{abstract}
With the increased use of virtual and augmented reality applications, the importance of point cloud data rises. High-quality capturing of point clouds is still expensive and thus, the need for point cloud super-resolution or point cloud upsampling techniques emerges. In this paper, we propose an interpolation scheme for color upsampling of three-dimensional color point clouds. As a point cloud represents an object's surface in three-dimensional space, we first conduct a local transform of the surface into a two-dimensional plane. Secondly, we propose to apply a novel Frequency-Selective Mesh-to-Mesh Resampling (FSMMR) technique for the interpolation of the points in 2D. FSMMR generates a model of weighted superpositions of basis functions on scattered points. This model is then evaluated for the final points in order to increase the resolution of the original point cloud. Evaluation shows that our approach outperforms common interpolation schemes.  Visual comparisons of the \textit{jaguar} point cloud underlines the quality of our upsampling results. The high performance of FSMMR holds for various sampling densities of the input point cloud.
\end{abstract}


\section{Introduction}
\label{Sec:introduction}
\begin{figure}[t]
\centering
\includegraphics[scale=.9, trim=145 75 105 20 mm, clip=true]{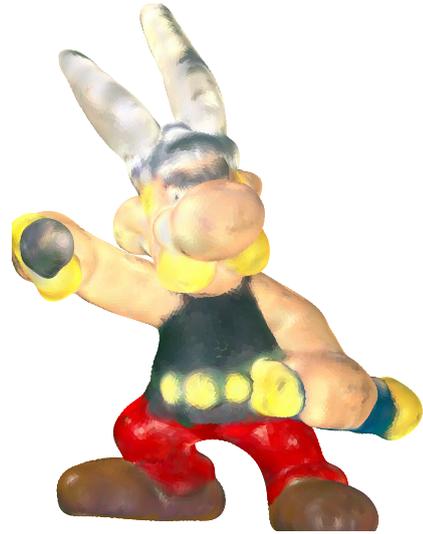}
\caption{\label{Fig:Asterix} \textit{Asterix} point cloud \cite{3DColMesh}.}
\end{figure}

Point clouds are an emerging three-dimensional data type and are often used in automated driving, virtual, and augmented reality. A point cloud is a set of three-dimensional coordinates with eventually assigned texture or color.  Point clouds mainly show object surfaces in three-dimensional space. The points are not restricted to lie on integer coordinates. Furthermore, the points are stored unordered, i.e., no connection to the points' neighborhood can be established from the stored point cloud.  An example of such a color point cloud is given in Figure \ref{Fig:Asterix}.\\
As the acquisition of point clouds is expensive, mainly low resolution point clouds are acquired. This is in contrast to the rendering for high resolution screens. Thus, super-resolution of point clouds becomes necessary.  Another term for point cloud super-resolution that is mainly driven from the computer science community is point cloud upsampling. We will use both interchangeable in the remainder of this paper.  Comparing super-resolution of point clouds to super-resolution of images, new challenges arise. As the points are allowed to obtain any coordinate in the three-dimensional space, first, the problem of geometry upsampling has to be solved.  Amenta et al.  \cite{Amenta} developed a common algorithm for reconstructing three-dimensional surfaces. Therefore, they compute a Voronoi diagram. In \cite{Alexa_2001},  Alexa et al. used point sets to represent shape. They add new points at the vertices of Voronoi diagrams that are built on moving least square surfaces. Data-driven approaches were developed in recent years with the rise of neural networks. Yu et al. developed PU-Net \cite{Yu_2018_PUNet}, the first network that upsamples point clouds. Its feature extraction is based on PointNet++ \cite{Qi_2017}, which classifies and segments point clouds.  Furthermore, the joint loss function aims to insert new points uniformly. In order to increase the performance of PU-Net especially for large resolution factors, the upsampling unit is applied recursively several times in MPU net \cite{Yifan_2019_MPU}. Li et al.  \cite{Li_2019_ICCV} presented another improvement of PU-Net with PU-GAN, a PU-Net that is embedded in a generator-discriminator structure in order to further increase the performance.With a focus on edge handling, Yu et al. presented EC-Net \cite{Yu_2018_EdgeAware}. It aims to better handle sharp edges in point cloud objects by inserting more points in edge-like areas. As these networks mainly work patch-based, Zhang et al.  \cite{Zhang_2019_DataDriven} presented a data-driven approach for the upsampling of point clouds by incorporating the entire point cloud and thus, focusing more on the overall shape of the object than on the local patch shape. \\
Super-resolution of color point clouds requires color upsampling in a second step. Therefore, common interpolation methods like linear, cubic, and natural neighbor interpolation are incorporated. Dinesh et al. \cite{Dinesh_2020} presented a graph-based approach that is applicable to both geometry and color upsampling. They create a k-nearest neighbor graph to estimate coordinates and RGB-values. These are then refined by minimizing a graph total variation.  \\
In this paper, we will focus on the color super-resolution of color point clouds. In the upcoming section, we propose a 3D to 2D coordinate transform and show our framework for color upsampling of color point clouds.  Then, we continue with the presentation of our novel frequency-selective mesh-to-mesh resampling algorithm. In Section \ref{Sec:Evaluation}, we show and interpret our results. Finally, the paper closes with a conclusion in Section \ref{Sec:Conclusion}.

\begin{figure}[t]
\centering
\scalebox{.9}{\input{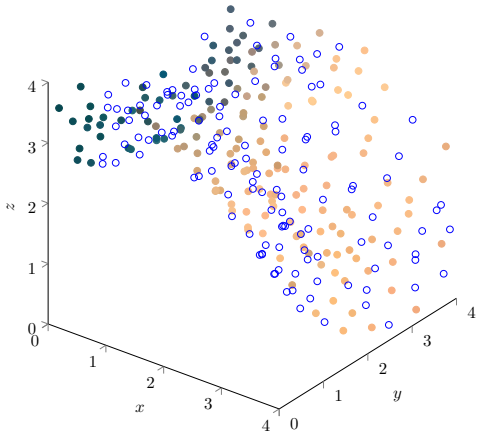}}
\caption{\label{Fig:asterix3} Block of the three-dimensional point cloud \textit{Asterix} shown in Figure \ref{Fig:Asterix} before color upsampling. Filled points show their respective color values (\textit{here}: varying from flesh-coloured to black and dark green). These colored points are the originally known points in point set $\mathcal{O}$. Color information has to be determined for the points denoted as blue circles belonging to the point set $\mathcal{R}$.  }
\end{figure}

\section{Color Upsampling}
\label{Sec:ColorUpsampling}
As described in Section \ref{Sec:introduction}, color point cloud super-resolution requires a geometry and color upsampling step. In this section, we focus on the color upsampling part. We assume that the geometry upsampling is already conducted applying one of the presented algorithms \cite{Amenta, Alexa_2001, Yu_2018_PUNet, Yifan_2019_MPU, Li_2019_ICCV, Yu_2018_EdgeAware, Zhang_2019_DataDriven, Dinesh_2020}.\\
For color upsampling, we assume to have a set of originally known points carrying coordinate and color information. We denote the set of original points as $\mathcal{O}$. Furthermore, we assume the geometry upsampling algorithms to generate a set of points carrying solely coordinates. Color information of these points is not known yet. We denote the set of the to be reconstructed points as $\mathcal{R}$.  The point cloud that we expect to input in our framework is the joint set  $\mathcal{P} = \mathcal{O} \cup \mathcal{R}$.  The point set $\mathcal{P}$ of a block of the \textit{Asterix} point cloud is given in Figure~\ref{Fig:asterix3}.  In the following, all computations are conducted locally on a block which is a $4\times 4\times 4$ cuboid from the three-dimensional point cloud. The colored and filled points are in $\mathcal{O}$ and show their respective RGB values.  The blue circles denote the points in $\mathcal{R}$. Color information is not yet known for these points. Their coordinates are generated using a geometry upsampling technique before.  We aim to reconstruct the color information of these points. On closer examination it becomes obvious that a point cloud represents an object's surface in three-dimensional space. As a surface is usually two-dimensional,we aim to transform the surface into a two-dimensional plane.  The transform works block-based and consists of five steps. 
First, the Euclidean distance $w_e$ is determined between the points of the joint set of point cloud coordinates $\mathcal{P}$
\begin{equation}
\label{Eq:euclidean}
\begin{split}
w_e(p_i, p_j) = \sqrt{(x_j-x_i)^2+(y_j-y_i)^2+(z_j-z_i)^2} \\
\forall (p_i, p_j) \in \mathcal{P}.
\end{split}
\end{equation} 

\begin{figure}
\centering
\scalebox{.9}{\input{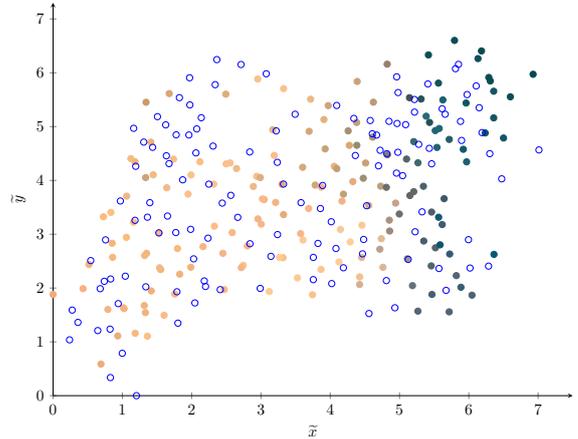}}
\caption{\label{Fig:asterix2} The same extract of the point cloud \textit{Asterix} as in Figure~\ref{Fig:asterix3} before color upsampling and after transforming to 2D. The filled and colored points are the originally known points in point set $\mathcal{O}$. Color information has to be determined for the points denoted as blue circles belonging to the point set $\mathcal{R}$.}
\end{figure}

\begin{figure}[t!]
\centering
%%% Copyright 2009 Jeffrey D. Hein
%%
%% This work may be distributed and/or modified under the
%% conditions of the LaTeX Project Public License, either version 1.3
%% of this license or (at your option) any later version.
%% The latest version of this license is in
%%   http://www.latex-project.org/lppl.txt
%% and version 1.3 or later is part of all distributions of LaTeX
%% version 2005/12/01 or later.
%%
%% This work has the LPPL maintenance status `maintained'.
%% 
%% The Current Maintainer of this work is Jeffrey D. Hein.
%%
%% This work consists of the files 3dplot.sty and 3dplot.tex
%
%%Description
%%-----------
%%3dplot.tex - an example file demonstrating the use of the 3dplot.sty package.
%
%%Created 2009-11-07 by Jeff Hein.  Last updated: 2009-11-09
%%----------------------------------------------------------
%
%%Update Notes
%%------------
%
%%2009-11-07: Created file along with 3dplot.sty package
%
%
%\documentclass{article}
%\usepackage{tikz}   %TikZ is required for this to work.  Make sure this exists before the next line
%
%\usepackage{3dplot} %requires 3dplot.sty to be in same directory, or in your LaTeX installation
%
%\usepackage[active,tightpage]{preview}  %generates a tightly fitting border around the work
%\PreviewEnvironment{tikzpicture}
%\setlength\PreviewBorder{2mm}
%
%\begin{document}
%
%%Angle Definitions
%%-----------------
%
%%set the plot display orientation
%%synatax: \tdplotsetdisplay{\theta_d}{\phi_d}
%\tdplotsetmaincoords{60}{110}
%
%%define polar coordinates for some vector
%%TODO: look into using 3d spherical coordinate system
\pgfmathsetmacro{\rvec}{.8}
\pgfmathsetmacro{\thetavec}{30}
\pgfmathsetmacro{\phivec}{60}

%start tikz picture, and use the tdplot_main_coords style to implement the display 
%coordinate transformation provided by 3dplot
\begin{tikzpicture}[scale=2]

%set up some coordinates 
%-----------------------
\coordinate (O) at (0,0,0);
\coordinate (P) at (0.05, 0.8, 0);
\coordinate (Q) at (0.4, 0.6, 0.1);
\coordinate (R) at (0.8, 0.4, 0.2);

%determine a coordinate (P) using (r,\theta,\phi) coordinates.  This command
%also determines (Pxy), (Pxz), and (Pyz): the xy-, xz-, and yz-projections
%of the point (P).
%syntax: \tdplotsetcoord{Coordinate name without parentheses}{r}{\theta}{\phi}
%\tdplotsetcoord{P}{\rvec}{\thetavec}{\phivec}
%\tdplotsetcoord{Q}{\rvec-.1}{\thetavec+10}{\phivec}
%\tdplotsetcoord{R}{\rvec+.1}{\thetavec}{\phivec-10}
%draw figure contents
%--------------------

%draw the main coordinate system axes
\draw[thick,->] (0,0,0) -- (1,0,0) node[anchor=north east]{$x$};
\draw[thick,->] (0,0,0) -- (0,1,0) node[anchor=north east]{$y$};
\draw[thick,->] (0,0,0) -- (0,0,1) node[anchor=south]{$z$};

% draw crosses at defined positions
%\draw[x] (P);
\draw (P) circle ;
\draw (Q) circle ;
\draw (R) circle ;

% fill circles with color
\fill [black]
(P) circle (0.03) node[above right] {$\{x_1, y_1, z_1\}$}
(Q) circle (0.03) node[right] {$\{x_2, y_2, z_2\}$}
(R) circle (0.03) node[below right] {$\{x_3, y_3, z_3\}$};

% draw conncection from P to Q and from Q to R
\draw[thick,color=black] (P) -- (Q);
\draw[thick,color=black] (Q) -- (R);

% draw arrow from 3D to 2D
\draw[thick, ->] (1.5,0.5) to (2,0.5) node[above left] {$\mathcal{T}$};

%% Add transformed coordinate system
%draw the main coordinate system axes
\draw[thick,->] (2.5,0) -- (3.5,0) node[anchor=north east]{$\widetilde{x}$};
\draw[thick,->] (2.5,0) -- (2.5,1) node[anchor=north east]{$\widetilde{y}$};
% define coordinates in 2D
\coordinate (P2) at (2.5, 0);
\coordinate (Q2) at (2.9,  1);
\coordinate (R2) at (3.3, 0.4);
% fill circles with color
\fill [black]
(P2) circle (0.03) node[below left] {$\{\widetilde{x}_1, \widetilde{y}_1\}$}
(Q2) circle (0.03) node[above] {$\{\widetilde{x}_2, \widetilde{y}_2\}$}
(R2) circle (0.03) node[below right] {$\{\widetilde{x}_3, \widetilde{y}_3\}$};
% draw conncection from P2 to Q2 and from Q2 to R2
\draw[color=black] (P2) -- (Q2);
\draw[color=black] (Q2) -- (R2);

\end{tikzpicture}

%\end{document}
\caption{\label{Fig:3to2transform}An extract of the minimum spanning tree is shown in the left image. The graph is transformed into 2D incorporating new coordinates $\widetilde{x}$ and $\widetilde{y}$ in the right graph. \vspace{-.2cm}}
\end{figure}
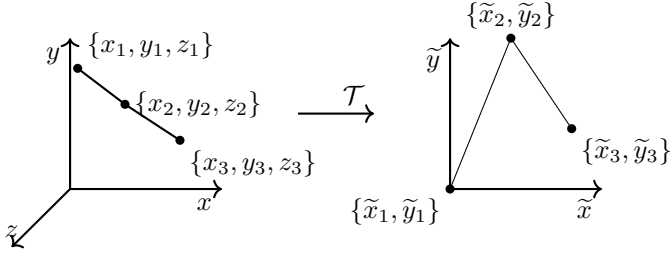

Knowing the euclidean distance between all points in the cloud, knowledge about neighborhood connections is established. Second, we further exploit this knowledge by creating a weighted graph $\mathcal{G} = \{\mathcal{V, E}\}$ with vertexes set $\mathcal{V}$ and edge set $\mathcal{E}$ with $(p_i, p_j) \in \mathcal{V}$ and  ${(p_i ,p_j, w_e(p_i, p_j))} \in\mathcal{E}$.  For the coordinate transform, we are interested in neighboring points. Thus, as a third step, a minimum spanning tree $\mathcal{M}$ is generated out of $\mathcal{G}$, i.e., $\mathcal{M}\subset \mathcal{G}$. Generating the minimum spanning tree enables us to calculate neighborhood relations.  An extract of a minimum spanning tree is given in Figure \ref{Fig:3to2transform}. Fourth, we exploit the established neighborhood relations. As we want to incorporate the third dimension into the first two dimensions, we calculate the Euclidean distance in first and third, and in second and third dimension for neighboring vertexes of the minimum spanning tree which are representing point cloud points yielding
\begin{equation}
\label{Eq:xtildedelta}
\Delta\widetilde{x}_{ij} = \text{sgn}(x_j-x_i) \sqrt{(x_j-x_i)^2+(z_j-z_i)^2}, 
\end{equation}
and
\begin{equation}
\label{Eq:ytildedelta}
\Delta\widetilde{y}_{ij} = \text{sgn}(y_j-y_i) \sqrt{(y_j-y_i)^2+(z_j-z_i)^2}.
\end{equation}
Using the $\text{sgn}$ function, we keep the direction between the points in x- and y-direction, respectively. Fifth, we add \eqref{Eq:xtildedelta} and \eqref{Eq:ytildedelta} to the previous vertex in the graph starting at the randomly selected root node from the minimum spanning tree. Thus, the transformed coordinates $\widetilde{x}_{j}$ and $\widetilde{y}_{j}$ are determined
\begin{equation}
\label{Eq:xtilde}
\widetilde{x}_{j} =\widetilde{x}_{i} + \Delta\widetilde{x}_{ij}
\end{equation}
\begin{equation}
\label{Eq:ytilde}
\widetilde{y}_{j} =\widetilde{y}_{i} + \Delta\widetilde{y}_{ij}.
\end{equation}
Thus, we map the 3D coordinates into 2D space and incorporate the third dimension into the new coordinates.  Thereby, we do not perform a projection that neglects the third dimension totally.  The 3D to 2D transform is summarized in Figure~\ref{Fig:3to2transform}. The exemplary extract of the \textit{Asterix} point cloud in Figure~\ref{Fig:asterix3} is shown after the transform in Figure~\ref{Fig:asterix2}.  The transform keeps neighboring points from 3D next to each other in 2D as well. The additionally inserted points for upsampling are located within the set of originally known color values. Thus, a two-dimensional mesh-to-mesh resampling problem is formulated. \\
The resampling problem can be solved by incorporating common interpolation schemes such as linear and cubic interpolation. In addition, we present a model-based approach called frequency-selective mesh-to-mesh resampling in the upcoming section. It aims to generate the missing color values $\mathcal{R}$. Finally, these have to be assigned to the according point in three-dimensional space. Hence, this yields the upsampled color point cloud.\\

\begin{figure}
\centering
% Define block styles

\definecolor{colorflow}{rgb}{0.00000,0.44700,0.74100}%

\tikzset{%
	back group/.style={fill=yellow!20,rounded corners, draw=black!50, dashed, inner xsep=12pt, inner ysep=11pt, yshift=-5pt}
}

\tikzstyle{decision} = [diamond, draw=colorflow, fill=white!15, line width=2pt,
    text width=4.5em, text badly centered, node distance=3cm, inner sep=0pt]
\tikzstyle{block} = [rectangle, draw, fill=white!15, 
    text width = 20em, text centered, rounded corners, minimum height=2em] % Standard Block
 \tikzstyle{block2} = [rectangle, draw=colorflow, fill=white!15, line width = 2pt,
    text width = 20em, text centered, rounded corners, minimum height=2em] % Standard Block blaue Umrandung für 2D
 \tikzstyle{blockk} = [rectangle, draw, fill=colorflow, 
    text width = 20em, text centered, rounded corners, minimum height=2em] % Block
\tikzstyle{blockkk} = [rectangle, draw=white, fill=white!15, 
    text width = 20em, text centered, rounded corners, minimum height=2em] % Block ohne Rahmen
\tikzstyle{blocksummarize} = [rectangle, draw, fill=white!15, line style=dashed,
    text width = 20em, text centered, rounded corners, minimum height=2em] % Block zum Zusammenfassen anderer Blöcke
\tikzstyle{line} = [draw, -latex']
\tikzstyle{cloud} = [draw, ellipse,fill=white!20, node distance=5cm,
    minimum height=2em]
    
\begin{tikzpicture}[node distance = 1.5cm, auto]
    % Place nodes
    \node [blockkk] (init) {3D low-resolution point cloud $\mathcal{O}$ and upsampled geometry points $\mathcal{R}$};
    \node [block, below of=init, node distance = 1.2cm] (euclidean) {Calculate Euclidean distance};
    \node [block, below of=euclidean] (graph) {Create graph $\mathcal{G} = \{\mathcal{V}, \mathcal{E}\}$};
    \node [block, below of=graph] (minspan) {Create minimum spanning tree $\mathcal{M} \subset \mathcal{G}$};
    \node [block2, below of=minspan] (deltatilde) {Calculate $\Delta\widetilde{x}_{ij}$ and $\Delta\widetilde{y}_{ij}$};
    \node [block2, below of=deltatilde] (tilde) {Calculate $\widetilde{x}_{j}$ and $\widetilde{y}_{j}$};
    \node [block2, below of=tilde,  node distance = 1.95cm] (residual) {Calculate residual $r^{(\nu)}[m,n]$};
    \node [block2, below of=residual] (energy) {Calculate residual energy decrease $\Delta E^{(\nu)}$ for every basis function};
    \node [block2, below of = energy] (selection) {Selection of best fitting basis function};
    \node [decision, below of=selection, node distance = 2.5cm] (stop) {Stopping criterium met?};
    \node [block2, below of= stop, node distance = 2.5cm] (mesh) {Obtain signal in $\mathcal{R}$};
    \node [block, below of=mesh, node distance = 1.6cm] (3to2) {Assign reconstructed values to $\{x, y, z\}\in \mathbb{R}^3$};
    \node [blockkk, below of=3to2, node distance = 1cm] (finish) {3D high-resolution color point cloud};
    
    % Create coordinates
    \coordinate[right of=residual] (a1);  %This is so the path doesn't intersect the diagram 
    \coordinate[right of=stop] (e1); 

    % Draw edges
    \path [line] (init) -- (euclidean);
    \path [line] (euclidean) -- (graph);
    \path [line] (graph) -- (minspan);
    \path [line] (minspan) -- (deltatilde);
    \path [line] (deltatilde) -- (tilde);
    \path [line] (tilde) -- (residual);
    \path [line] (residual)  -- (energy);
    \path [line] (energy) -- (selection);
    \path [line] (selection) -- (stop);
    \path [line] (stop) -| node [near start]{No}([xshift=3.0cm]e1) -- node[sloped, anchor=center, above, text width = 5cm](atest){Generated model $g^{(\nu)}[m,n]$}([xshift=3.0cm]a1) -- (residual);
    \path [line] (stop) -- node[near start]{Yes} (mesh);
    \path [line] (mesh) -- (3to2);
    \path [line] (3to2) -- (finish);
    
    \path (atest) +(0.3, 10) coordinate (a2);
    
    \begin{scope}[on background layer]
    		\node (bk1) [back group] [fit=(euclidean) (a2) (tilde)] {};
    		\node (bk2) [back group] [fit=(residual) (atest) (mesh)] {};
    		\node[anchor=south east,inner sep=1pt,outer sep=1pt,opacity=0.5,font=\sffamily\bfseries,text=gray] at (bk1.south east){3D to 2D transform};
    		\node[anchor=south east,inner sep=1pt,outer sep=1pt,opacity=0.5,font=\sffamily\bfseries,text=gray] at (bk2.south east){FSMMR};
    \end{scope}
    
%\node[fill=orange,rounded corners=1cm,text width=0.2\textwidth,inner sep=0.4cm] {\lipsum[2]};
%\draw[dotted] (2.2,6.5) rectangle (7.8,1.3) node at (6.5,1.45) []{Data Augmentation};    

% following \eqref{Eq:evaluationonmesh}};
    % -- Sec. \ref{Sec:ColorUpsampling}};
    % -- Sec. \ref{Sec:FSMMR}};
\end{tikzpicture}
 \vspace{-.2cm}
\caption{\label{Fig:flowgraph}Summary of the proposed framework. 3D low-resolution point cloud is inserted into the framework. The first yellow shaded block represents the 3D to 2D transform. The second one summarizes FSMMR. Thick, blue blocks compute 2D signals. The final output is the 3D high-resolution point cloud.}
\end{figure}
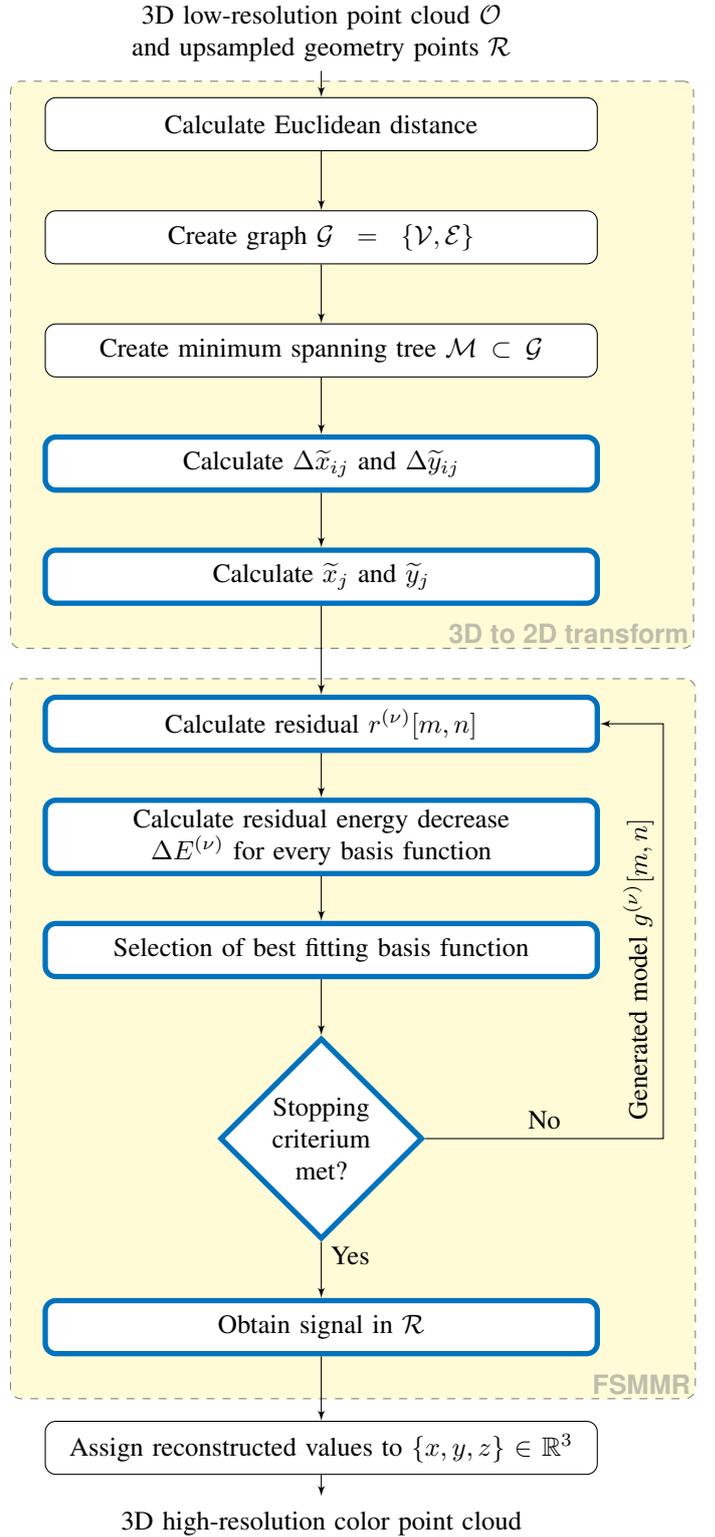

 \vspace{-.3cm}
\section{Frequency-Selective Mesh-to-Mesh Resampling}
\label{Sec:FSMMR}
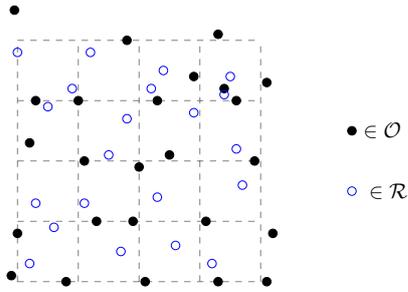
\begin{figure}
\centering
\scalebox{.8}{\begin{tikzpicture}
\tkzInit[xmax=4, ymax=4]
\begin{scope}[dashed]
\tkzGrid
\end{scope}
\draw[fill=black](-0.1,0.1)circle(2pt);
\draw[fill=black](0,0.8)circle(2pt);
\draw[fill=black](0.2,2.3)circle(2pt);
\draw[fill=black](0.3,3)circle(2pt);	
\draw[fill=black](-0.05,4.5)circle(2pt);	
\draw[fill=black](0.8,0)circle(2pt);
\draw[fill=black](1.3,1)circle(2pt);	
\draw[fill=black](1.1,2)circle(2pt);
\draw[fill=black](1,3)circle(2pt);	
\draw[fill=black](1.8,4)circle(2pt);
\draw[fill=black](2.1,0)circle(2pt);
\draw[fill=black](2,1.9)circle(2pt);
\draw[fill=black](2.5,2.1)circle(2pt);
\draw[fill=black](2.3,3)circle(2pt);	
\draw[fill=black](2.9,3.4)circle(2pt);	
\draw[fill=black](3.3,0)circle(2pt);
\draw[fill=black](3.1,1)circle(2pt);	
\draw[fill=black](3.9,2)circle(2pt);
\draw[fill=black](3.6,3)circle(2pt);	
\draw[fill=black](3.3,4.1)circle(2pt);
\draw[fill=black](4.1,0)circle(2pt);
\draw[fill=black](1.9,1)circle(2pt);	
\draw[fill=black](4.2,0.8)circle(2pt);
\draw[fill=black](4.1,3.3)circle(2pt);	
\draw[fill=black](3.4,3.2)circle(2pt);	
\draw[color=blue](0.2, 0.3)circle(2pt);
\draw[color=blue](0.3, 1.3)circle(2pt);
\draw[color=blue](0.6, 0.9)circle(2pt);
\draw[color=blue](0.5, 2.9)circle(2pt);
\draw[color=blue](0.9, 3.2)circle(2pt);
\draw[color=blue](0, 3.8)circle(2pt);
\draw[color=blue](1.2, 3.8)circle(2pt);
\draw[color=blue](1.7, 0.5)circle(2pt);
\draw[color=blue](1.5, 2.1)circle(2pt);
\draw[color=blue](1.1, 1.3)circle(2pt);
\draw[color=blue](1.8, 2.7)circle(2pt);
\draw[color=blue](2.6, 0.6)circle(2pt);
\draw[color=blue](2.3, 1.4)circle(2pt);
\draw[color=blue](2.4, 3.5)circle(2pt);
\draw[color=blue](2.9, 2.8)circle(2pt);
\draw[color=blue](2.2, 3.2)circle(2pt);
\draw[color=blue](3.2, 0.3)circle(2pt);
\draw[color=blue](3.7, 1.6)circle(2pt);
\draw[color=blue](3.5, 3.4)circle(2pt);
\draw[color=blue](3.6, 2.2)circle(2pt);
\draw[color=blue](3.4, 3.1)circle(2pt);

\draw[fill=black](5.5, 2.5)circle(2pt);
	\draw[color=blue](5.5, 1.5)circle(2pt);
	\node at (6, 2.5) {$\in \mathcal{O}$};
	\node at (6.1, 1.5) {$\in \mathcal{R}$};
\end{tikzpicture}}
\caption{\label{Fig:mesh2mesh}The coordinates of the original point set $\mathcal{O}$ as black dots and the to be reconstructed point set $\mathcal{R}$ as blue circles are shown. \vspace{-.4cm}}
\end{figure}

In this section, we introduce the novel Frequency-Selective Mesh-to-Mesh Resampling (FSMMR) for color upsampling of point clouds.  As described in Section \ref{Sec:ColorUpsampling} color upsampling of point clouds requires an interpolation that can create new color values at arbitrary points, which we will refer to as mesh in the following, from original points at other mesh positions.  We refer to the set of original mesh points as $\mathcal{O}$ and to the set of to be reconstructed points as $\mathcal{R}$. Both sets are shown exemplary in Figure \ref{Fig:mesh2mesh}.\\
Research has shown that image signals can be represented in terms of weighted superpositions of basis functions. This assumption is used for the extrapolation into unknown image areas in Frequency-Selective Extrapolation \cite{2005_Kaup_FSEOrig}, to solve irregular sampling problems using Frequency-Selective Reconstruction \cite{Schoeberl_2011}, and for resampling from originally known mesh positions onto unknown grid points in Frequency-Selective Mesh-to-Grid Resampling \cite{2017_Koloda_FSMR, Heimann_2020_MMSP}. For color upsampling of point clouds, the surface of a three-dimensional object has to be upsampled. We assume, that the two-dimensional surface $f[m,n]$ of a three-dimensional point cloud can also be represented as superposition of weighted basis functions $\varphi_{(k, l)}$
\begin{equation}
\label{Eq:image}
f[m, n] = \sum_{k, l \in \mathcal{K}} c_{k, l} \varphi_{k, l}[m, n],
\end{equation}   
where $m \in \mathbb{R}\cup\mathcal{O}$ and $n \in \mathbb{R}\cup\mathcal{O}$ denote the points' arbitrary coordinates in horizontal and vertical direction, respectively. The indexes $k \in \mathbb{N}$ and $l \in \mathbb{N}$ denote the frequency indexes in horizontal and vertical direction, respectively, from the set of available basis functions $\mathcal{K}$. The assigned expansion coefficient $c$ can be interpreted as transform coefficient of the inverse transform as basis functions from the discrete cosine transform (DCT) are incorporated.  The aim in FSMMR is to create a model $g[m,n]$ that approaches \eqref{Eq:image}. The model is generated block-wise and iteratively on the set of original points $\mathcal{O}$
\begin{equation}
\label{Eq:modelGeneration}
g^{(\nu)}[m, n] = g^{(\nu -1)}[m, n] + \hat{c}_{u, v} \varphi_{u, v}[m, n],
\end{equation}
where $\nu$ denotes the current iteration number and $\hat{c}_{u, v}$ the estimated expansion coefficient of the selected frequency coefficients $u \in \mathbb{N}$ and $v \in \mathbb{N}$ in the current iteration.  The model is initialized to zero, i.e. $g^{(0)}\equiv 0$. The model's aim is to meet the original signal $f[m,n]$ as close as possible. Thus, the residual between both is determined 
\begin{equation}
\label{Eq:Residual}
r^{(\nu)} = f[m,n] - g^{(\nu)} [m,n].
\end{equation}
As the residual has to be reduced, a weighted residual energy is defined
\begin{equation}
E^{(\nu)} = \sum_{(m,n)} w[m,n]\left(r^{(\nu)}[m,n]\right)^2.
\end{equation}
with a spatial weighting function $w[m,n]$ that favors center points for the model generation.  Furthermore, research has shown that natural images are mainly composed of low frequency basis functions  \cite{2015_Seiler_ResamplingImages}. We assume that this holds for objects' surfaces as well. Hence, we incorporate the frequency weighting function 
\begin{equation}
\label{Eq:FreqWeighting}
w_{f}[k,l] = \sigma^{\sqrt{k^2 + l^2}},
\end{equation}
where $\sigma \in ]0,1[$ parametrizes the decay, into the selection of the best fitting basis function in iteration~$\nu$
\begin{equation}
{(u, v)} = \underset{{(k, l)}}{\mathrm{argmax}} \left( \Delta E_{k, l}^{(\nu)} w_{f}[k,l] \right).
\end{equation}
In every iteration, the basis function is selected that maximizes the residual energy reduction as it will close the gap between model and original signal the most. These steps are repeated until a stopping criterion such as a number of iterations or a minimal residual energy is met. \\
Finally, the model is evaluated for the reconstructed points in $\mathcal{R}$. Thus, the generated estimated expansion coefficients $\hat{c}_{(k, l)}$ are multiplied by the basis functions $\varphi_{(k, l)}$
\begin{equation}
\label{Eq:evaluationonmesh}
f[o,p] = \sum_{k, l \in \mathcal{K}} \hat{c}_{k, l} \varphi_{k, l}[o, p],
\end{equation}   
at the newly inserted mesh points $[o, p] \in \mathbb{R}^2\in\mathcal{R}$.\\
A summary of the proposed color upsampling scheme incorporating FSMMR is shown as a flow graph in Figure \ref{Fig:flowgraph}.

\section{Experimental Results}
\label{Sec:Evaluation}

\begin{figure}[t]
\definecolor{mycolor1}{rgb}{0.00000,0.44700,0.74100}%
\definecolor{mycolor2}{rgb}{0.85000,0.32500,0.09800}%
\definecolor{mycolor3}{rgb}{0.92900,0.69400,0.12500}%
\definecolor{mycolor4}{rgb}{0.49400,0.18400,0.55600}%
\definecolor{mycolor5}{rgb}{0.46600,0.67400,0.18800}%
\begin{tikzpicture}

\begin{axis}[%
width=.35\textwidth,
height=.25\textwidth,
at={(1.95in,0.85in)},
scale only axis,
xmin=10,
xmax=80,
ymin=8,
ymax=26,
xtick distance = 10,
ytick distance = 2.5,
ylabel = {PSNR in dB},
xlabel = {Sampling density in \%},
axis background/.style={fill=white},
every axis plot/.append style={very thick},
title style={font=\bfseries},
legend style={legend style={at={(.175\textwidth, .27\textwidth)},anchor=south}, legend columns=3, legend cell align=left, align=left, draw=white!15!black}
]

\addplot [color=mycolor1, style=dashed]
  table[row sep=crcr]{%
10	19.368073491575\\
20	20.6462573429307\\
30	21.381513186926\\
40	21.8759820324104\\
50	22.2106332841961\\
60	22.5704240632372\\
70	22.7250324777549\\
80	22.9529005034702\\
};
\addlegendentry{LIN3}

\addplot [color=mycolor3, style=dashed]
  table[row sep=crcr]{%
10	19.4214999710204\\
20	20.6965235570126\\
30	21.4289229326313\\
40	21.9259786773558\\
50	22.2624113970325\\
60	22.6262511430514\\
70	22.784516975369\\
80	23.01462771718\\
};
\addlegendentry{NAT3}

\addplot [color=mycolor1, line width = 3pt]
  table[row sep=crcr]{%
10	9.48931813666191\\
20	10.9176928811147\\
30	11.7236768830091\\
40	12.4253444210601\\
50	13.0545324316617\\
60	13.5313626621921\\
70	13.905707682706\\
80	14.2380893587863\\
};
\addlegendentry{LIN2}

\addplot [color=mycolor2]
  table[row sep=crcr]{%
10	9.45968283451316\\
20	10.8836021429851\\
30	11.6861458366135\\
40	12.3820081144813\\
50	13.0074252351191\\
60	13.4788919403017\\
70	13.8489994775048\\
80	14.1762078305008\\
};
\addlegendentry{CUB2}

\addplot [color=mycolor3]
  table[row sep=crcr]{%
10	9.51020495489423\\
20	10.9438520680002\\
30	11.7504025196219\\
40	12.4594714894697\\
50	13.0887828050991\\
60	13.5679706727573\\
70	13.943445671496\\
80	14.2802253529428\\
};
\addlegendentry{NAT2}

\addplot [color=mycolor5]
  table[row sep=crcr]{%
10	19.9783980649267\\
20	21.5940145943095\\
30	22.6056065056015\\
40	23.287561481935\\
50	23.7400342114053\\
60	24.0798633902396\\
70	24.2822793136109\\
80	24.4650440332748\\
};
\addlegendentry{FSMMR}

\end{axis}
\end{tikzpicture}%
\caption{\label{Fig:samplingdens} Reconstrution PSNR in terms of dB is given on average for the entire \textit{3D Color Mesh} dataset dependent on the sampling density of the input point cloud, i.e., the enlargement factor of the point cloud.  3D approaches are given in dashed lines. The best performing method is FSMMR for all sampling densities.}\vspace{-.5cm}
\end{figure}
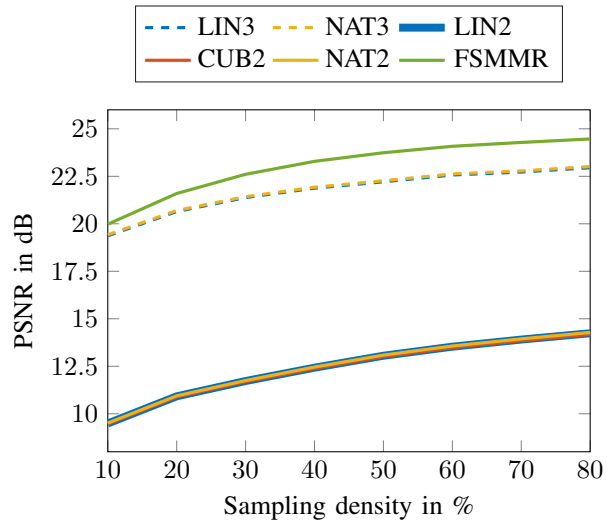
\begin{figure*}
\centering
\begin{subfigure}{.25\textwidth}
\centering
 {\includegraphics[scale=1.1, trim=140 75 190 80 mm, clip=true]{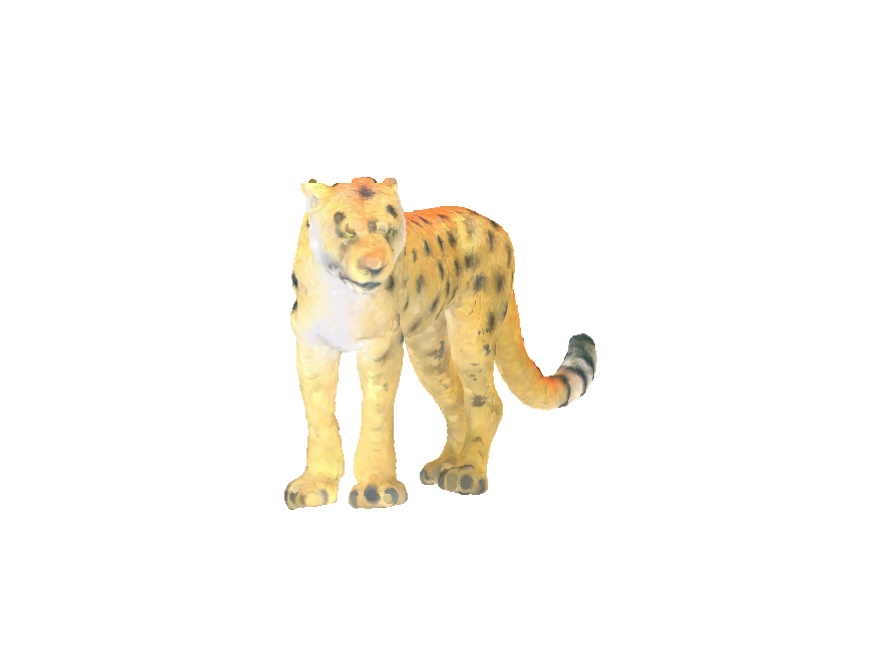}} % trim = left bottom right top
\caption{\label{Fig:jaguar_orig}Original.}
\end{subfigure}\hfill%
\begin{subfigure}{.25\textwidth}
\centering
 {\includegraphics[scale=1.18, trim=145 75 190 90 mm, clip=true]{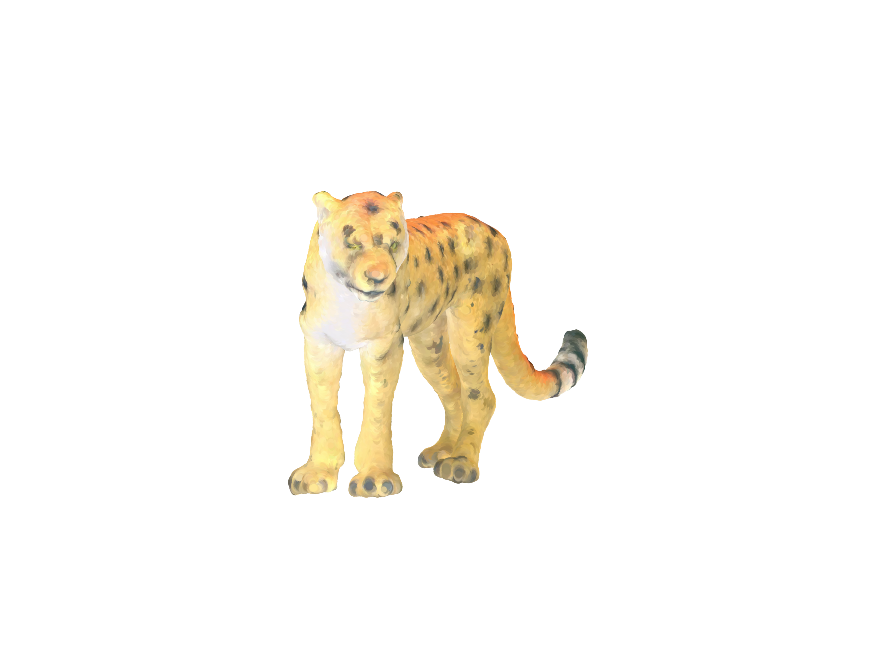} }% trim = left bottom right top
\caption{\label{Fig:jaguar_downsampled}Low-resolution.}
\end{subfigure}\hfill%
\begin{subfigure}{.25\textwidth}
\centering
 {\includegraphics[scale=.8, trim=115 35 185 60 mm, clip=true]{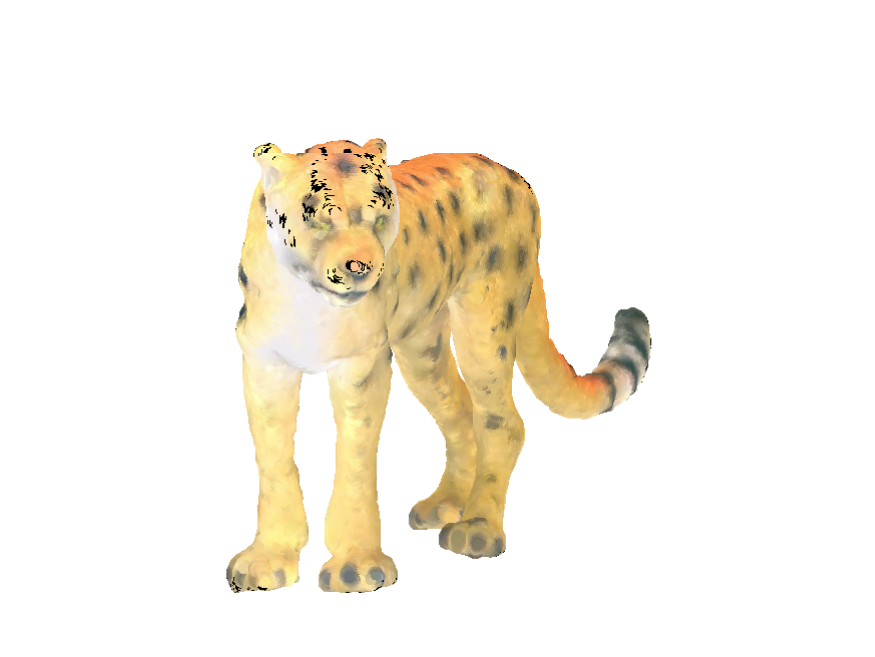}} % trim = left bottom right top
\caption{\label{Fig:jaguar_lin3}LIN3.}
\end{subfigure}\hfill%
\begin{subfigure}{.25\textwidth}
\centering
 {\includegraphics[scale=.8, trim=115 35 185 60 mm, clip=true]{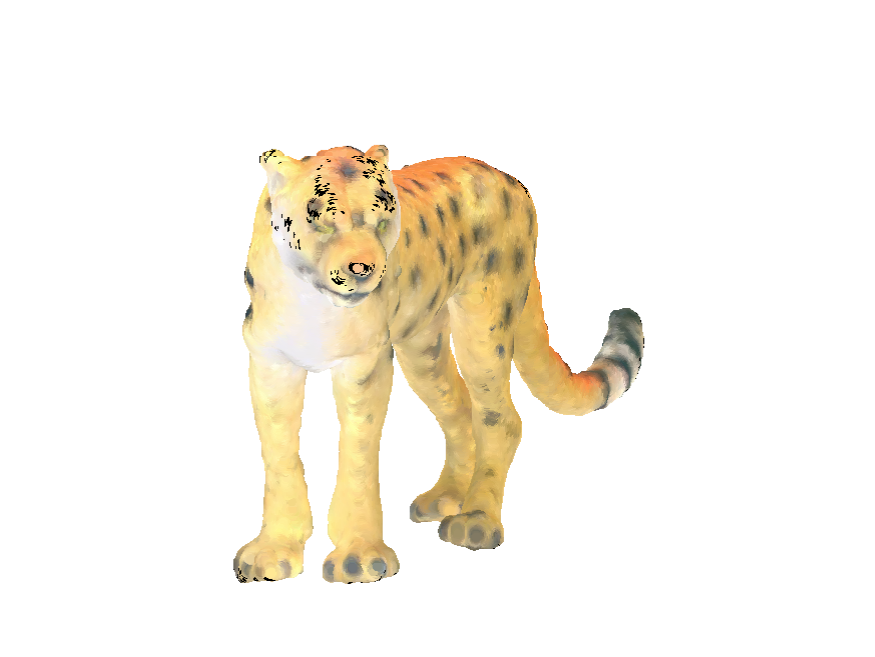}} % trim = left bottom right top
\caption{\label{Fig:jaguar_nat3}NAT3.}
\end{subfigure} \\
\centering
\begin{subfigure}{.25\textwidth}
\centering
 {\includegraphics[scale=.8, trim=115 35 185 60 mm, clip=true]{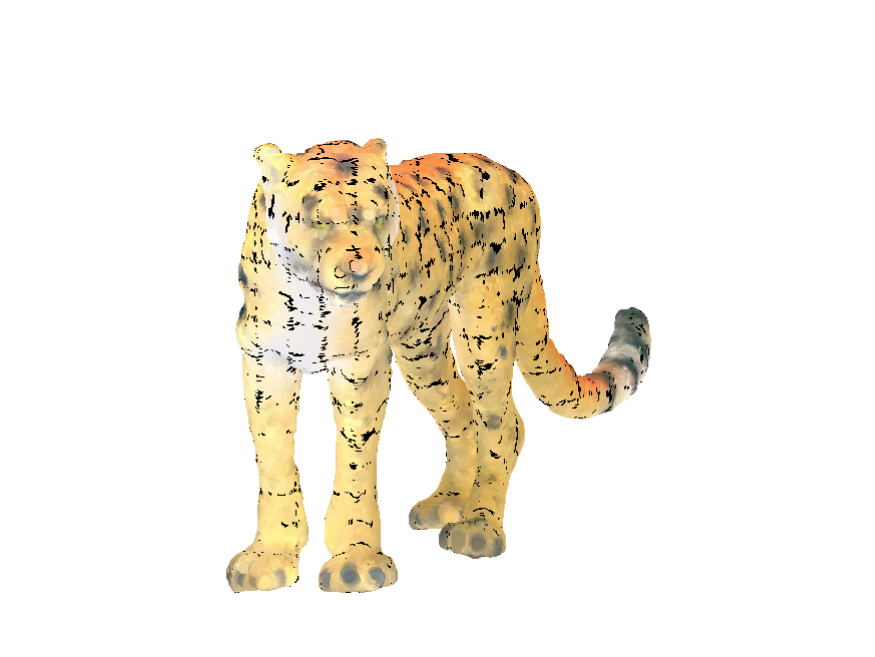}} % trim = left bottom right top
\caption{\label{Fig:jaguar_lin2}LIN2.}
\end{subfigure}\hfill%
\begin{subfigure}{.25\textwidth}
\centering
 {\includegraphics[scale=.8, trim=115 35 185 60   mm, clip=true]{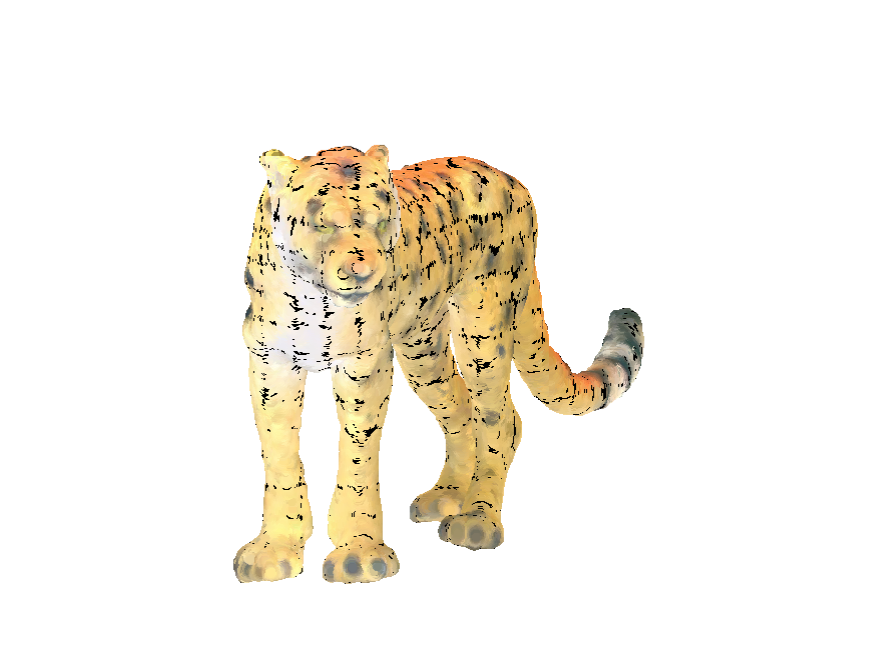}} % trim = left bottom right top
\caption{\label{Fig:jaguar_cub2}CUB2.}
\end{subfigure}\hfill%
\begin{subfigure}{.25\textwidth}
\centering
 {\includegraphics[scale=.8, trim=115 35 185 60   mm, clip=true]{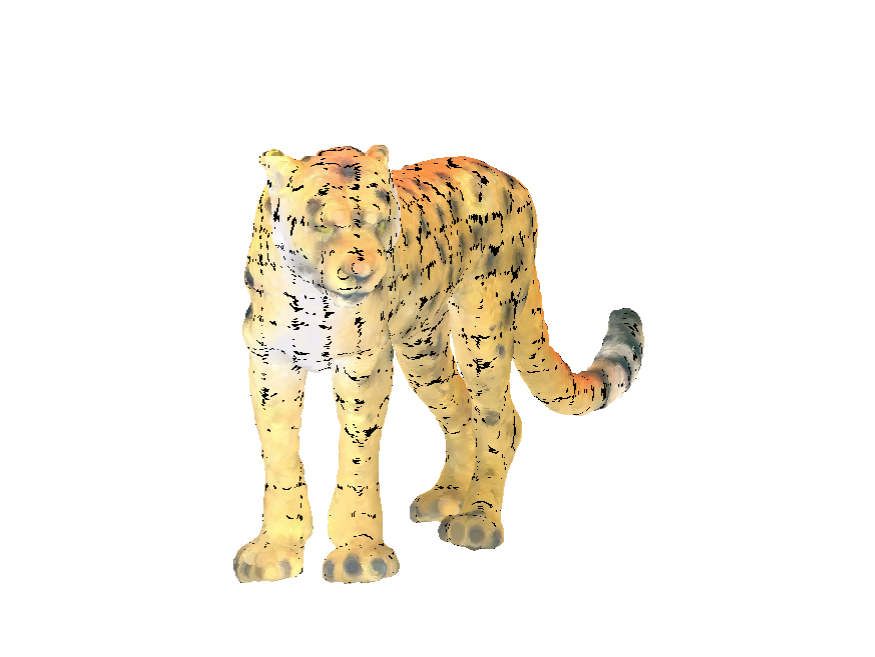}} % trim = left bottom right top
\caption{\label{Fig:jaguar_nat2}NAT2.}
\end{subfigure}\hfill%
\begin{subfigure}{.25\textwidth}
\centering
 {\includegraphics[scale=.8, trim=115 35 185 60   mm, clip=true]{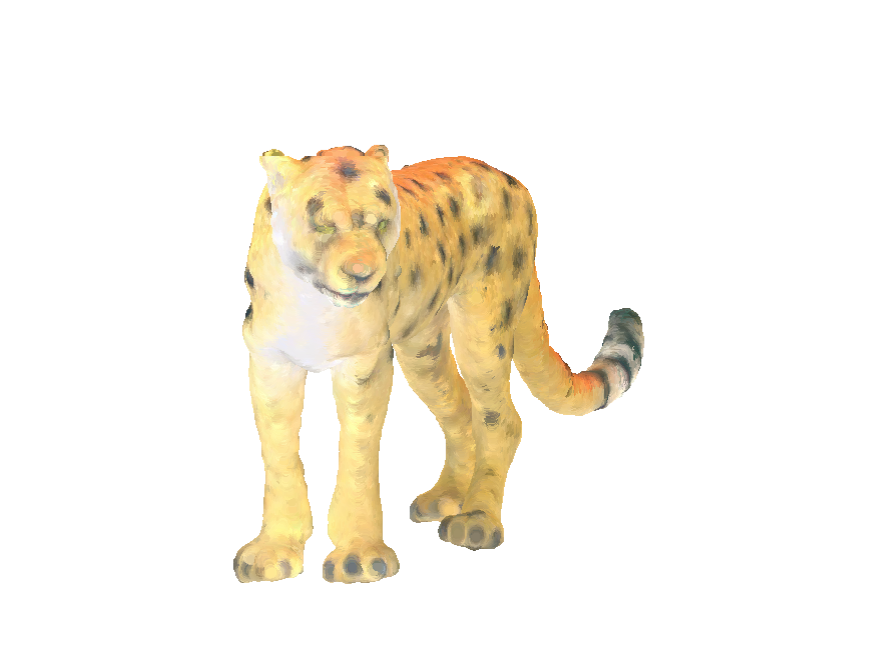} }% trim = left bottom right top
\caption{\label{Fig:jaguar_afsmr}FSMMR.}
\end{subfigure}\hfill%
\caption{\label{Fig:jag}The \textit{jaguar} point cloud. Best viewed enlarged on a screen.}\vspace{-.5cm}
\end{figure*}

We conduct extensive experiments in order to show the performance of our proposed color upsampling scheme.  We denote our proposed geometry transform from three-dimensional to two-dimensional surface as \textit{2D} approach described in Section \ref{Sec:ColorUpsampling}.  For the upcoming interpolation, we compare linear (LIN2), cubic (CUB2), natural neighbor interpolation (NAT2), and our proposed FSMMR. Furthermore, we evaluate the reconstruction quality without the proposed transform and apply linear (LIN3) and natural neighbor (NAT3) interpolation directly to the three-dimensional point cloud. Thus, we refer to this approach as \textit{3D}. \\
In order to evaluate the quality of the reconstructed point clouds, we first downsample the original ones randomly. The selected points are taken as low-resolution point cloud and are handed over to the proposed algorithm as original point set $\mathcal{O}$.  As we focus on color upsampling in this paper, the coordinates of the skipped points are kept in order to reconstruct the missing color values at the right positions. These points form the point set $\mathcal{R}$. This approach enables us to evaluate the final result in terms of peak signal-to-noise ratio (PSNR). We measure color PSNR conducting two steps. First, the PSNR is determined for the three color channels R, G, and B separately. Second, the average of the independently calculated PSNR values on each color channel is taken as color PSNR. We conducted every experiment for three runs in order to exclude effects that might occur due to the random selection of low-resolution points. All results that we show in the following are averaged values for the three runs. In one run, every interpolation scheme receives the same downsampled point cloud. Thus, the presented results are fully comparable.\\
The reconstruction quality of the point clouds of the \textit{3D Color Mesh} dataset \cite{3DColMesh} for a sampling ratio of $50\%$ is shown in Table \ref{Tab:Res3D} in terms of reconstruction PSNR in dB. The reconstruction PSNR measures the PSNR solely for the reconstructed points. Original points are not incorporated in the quality measurement. Best performances are highlighted in bold font. Table \ref{Tab:Res3D} demonstrates that our proposed approach incorporating FSMMR performs best for most point clouds. If FSMMR is not the best performing method such as for the \textit{CableCar} point cloud, the three-dimensional approaches, LIN3 and NAT3, yield the highest reconstruction quality. The two-dimensional approaches using common interpolation schemes never yield best result. Thus, the transform from 3D to 2D can be neglected as reason for the high quality of our proposed FSMMR scheme.  \\
Furthermore, we evaluated the performance of the \textit{3D Color Mesh} dataset for various sampling densities. A sampling density of $10\%$ means that only $10\%$ of the original points are kept and used for the reconstruction of the point cloud. Thereby, we simulate various enlargement factors of the point cloud. Thus, it is justified to expect smaller PSNR values for smaller sampling densities as more points have to be reconstructed based on less data. The averaged reconstruction PSNR for sampling densities from $10\%$ to $80\%$ is given in Figure~\ref{Fig:samplingdens}. The PSNR is averaged for the entire \textit{3D Color Mesh} dataset. As expected, the higher the sampling density, the better the reconstruction result for the point clouds. Moreover, the figure demonstrates that FSMMR is the best performing technique for all sampling densities.\\
A visual example for the reconstruction quality is given for the \textit{Jaguar} point cloud in Figure~\ref{Fig:jag}. In Figure~\ref{Fig:jaguar_orig}, the original point cloud is presented. In Figure~\ref{Fig:jaguar_downsampled}, the low-resolution point cloud which is the starting point for the upsampling algorithms is shown. In Figure~\ref{Fig:jaguar_lin3} and~\ref{Fig:jaguar_nat3}, the final point cloud is shown for LIN3 and NAT3 methods, respectively. Both point clouds show missing values in the region of the head of the jaguar, especially the left ear, the nose and the region above the eyes are affected by that.  This effect originates in the definition of linear and natural neighbor interpolation. In the border regions pixel values have to be extrapolated. A classical interpolation scheme cannot fullfill this requirement. The 2D approaches LIN2, CUB2, and NAT2 in Figure~\ref{Fig:jaguar_lin2}, ~\ref{Fig:jaguar_cub2}, and~\ref{Fig:jaguar_nat2}, respectively, show missing values along the block borders that are used for the proposed 3D to 2D transform.  The number of missing pixels increases due to the growing number of borders as a boarder occurs for every block. The result for our proposed FSMMR approach is shown in Figure~\ref{Fig:jaguar_afsmr}. No missing pixels can be detected, nor at the head of the jaguar nor at block borders within the point cloud although the same block structures are used for FSMMR as well as for LIN2, CUB2, and NAT2. This demonstrates that FSMMR can conduct interpolation and extrapolation in one step. A slight color shift into green can be detected at the mouth of the jaguar for FSMMR.

\begin{table}
\caption{\label{Tab:Res3D}Results for all point clouds from the \textit{3D Color Mesh} dataset in terms of reconstruction PSNR in dB. Best qualities are given in bold.}
\begin{tabular}{|l|c|c|c|c|c|c|}
\hline
							& \multicolumn{2}{c}{3D} & \multicolumn{4}{|c|}{2D} \\
\hline
Point Cloud			& LIN3 & NAT3 & LIN2 & CUB2 & NAT2  & FSMMR\\
\hline
\hline
 4armsMonstre 	& 23.4	 & 23.5 & 14.1	 & 14.9	 & 14.9  & \textbf{27.2}\\
\hline
 Asterix 				& 21.2 & 	21.2	 & 12.2	 & 12.2	 & 12.3	 & \textbf{23.0}\\
\hline
CableCar				& 24.5	 & \textbf{24.6}	 & 13.6	 & 13.6	 & 13.7	  & 22.3\\
\hline
 Dragon				& 26.7	 & 26.7 & 	15.5	 & 15.5	 & 15.5	& \textbf{27.7}\\
\hline
Duck						& 13.6  & 	13.7	 & 4.3 &  4.3	 & 4.3 & \textbf{15.2}\\
\hline
GreenDinosaur	& 25.3	 & 25.4	 & 15.3	 & 15.2 & 	15.3   & \textbf{25.6}\\
\hline
GreenMonstre		& 26.5	 & \textbf{26.7}	 & 16.6	 & 16.6	 & 16.6	 & 26.2\\
\hline
Horse					& 23.6	 & \textbf{23.7}	 & 10.5 & 	10.5	& 10.6	 & 20.3\\
\hline
 Jaguar					& 21.5	 & 21.5	 & 12.8	 & 12.8	 & 12.8	 & \textbf{27.0}\\
\hline
LongDinosaur 		& 21.5  & 21.5	 &  12.1	 & 12.1	 & 12.1	 & \textbf{28.8}\\
\hline
Man						& 30.3	 & \textbf{30.5}	 &  24.7	 & 24.2	 & 25.1	 & 20.7\\
\hline
Mario					& 23.4	 & 23.5 & 	14.7	 & 14.7	 & 14.7	  &  \textbf{24.5}\\
\hline
MarioCar				& 23.9  & 	23.9	 & 16.0	 & 16. 0 & 	16.1	  & \textbf{24.7}\\
\hline
PokemonBall		& 8.9	 & 8.9	 &  7.7	 & 7.7	& 7.7		 & \textbf{20.5}\\
\hline
Rabbit					& 21.0	 & 21.0 & 	10.2	 & 10.2	 & 10.2	 & \textbf{24.3}\\
\hline
RedHorse			& \textbf{22.9} & 	\textbf{22.9} & 	11.5	 & 11.5	 & 11.6	 & 22.1\\
\hline
Statue					& 24.3 &  24.3 & 	14.2	 & 14.2	 & 14.3	& \textbf{25.9}\\
\hline
\end{tabular}
\vspace{-.3cm}
\end{table}

\vspace{-.3cm}
\section{Conclusion}
\label{Sec:Conclusion}
In conclusion, we propose to use the frequency-selective mesh-to-mesh resampling technique for color upsampling of 3D color point clouds.  As we assume the point cloud to represent a surface,  we transform it into two-dimensional space. The transform enables us to apply FSMMR for color upsampling of 3D point clouds properly. The extensive evaluation shows that our proposed method works best for all sampling densities, i.e., enlargement factors. The visual comparison shows that only slight color shifts can occur using FSMMR, but all desired points are reconstructed and no missing pixels occur in border regions of blocks.

\section*{Acknowledgment}
The authors gratefully acknowledge that this work has been supported by the Deutsche Forschungsgemeinschaft (DFG) under contract number KA 926/8-1.
 \vspace{-.05cm}
\bibliographystyle{IEEEbib}
\bibliography{bib_4mmsp2021.bib}
\end{document}